\documentclass[aps,twocolumn,pra]{revtex4-2}
\usepackage{graphics}
\usepackage{helvet}
\usepackage{fancyhdr}
\usepackage{dcolumn}% Align table columns on decimal point
\usepackage{bm}% bold math
\usepackage{amsmath}
\usepackage{hyperref}
\hypersetup{colorlinks=true, urlcolor=blue,citecolor=blue,linkcolor=red}
\usepackage{color}% color to the text 
\usepackage{amssymb}
\usepackage{amsfonts}
\usepackage{graphicx}
\usepackage{physics}
\usepackage{amsthm}
\usepackage{float}
\usepackage{subfigure}

\newcommand{\orcid}[1]{\href{https://orcid.org/#1}{\resizebox{10px}{!}{\includegraphics{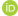}}}}
\begin{document}
%\allowdisplaybreaks % To allow page breaks in display
%\addtolength{\jot}{1em} % To allow for more vertical spacing in the align environment
% You should use BibTeX and revtex.bst for references

\title{\bf Multipath wave-particle duality with a path detector in a quantum superposition}

\author{Mohd Asad Siddiqui\orcid{0000-0001-5003-7571}}
\email{asad@ctp-jamia.res.in}
\affiliation{Institute of Fundamental and Frontier Sciences, University of Electronic Science and
	Technology of China, Chengdu 610051, China.}

\author{Tabish Qureshi\orcid{0000-0002-8452-1078}}
\email{tabish@ctp-jamia.res.in}
\affiliation{Centre for Theoretical Physics, Jamia Millia Islamia, New Delhi 110025, India.}

%\date{\today}

\begin{abstract}
According to Bohr's principle of complementarity, a quanton can behave either as a wave or a particle, depending on the choice of the experimental setup. Some recent two-path interference experiments have devised methods where one can have a quantum superposition of the two choices, thus indicating that a quanton may be in a superposition of wave and particle natures. These experiments have been of interest from the point of view of Wheeler's delayed-choice experiment. However, it has also been claimed that this experiment can violate complementarity. Here we theoretically analyze a multipath interference experiment that has a which-path detector in a quantum superposition of being present and absent. We show that a tight multipath wave-particle duality relation is respected in all such situations, and complementarity holds well. The apparent violation of complementarity may be due to incorrect evaluation of  path distinguishability in such scenarios.
\vskip 2mm
\noindent DOI: \href{https://doi.org/10.1103/PhysRevA.103.022219}{10.1103/PhysRevA.103.022219}
\end{abstract}
%\pacs{03.65.Bz,89.70.+c, 03.67.Mn, 42.50.Dv }

\maketitle
\thispagestyle{fancy}
\chead{PHYSICAL REVIEW A \textbf{103}, 022219 (2021)}

\section{Introduction}
 The discourse of wave-particle duality has always attracted attention from the early days of quantum mechanics. It is believed that it lies at the heart of quantum mechanics \cite{feynman}. It was understood from the beginning that the object exhibits both wave and particle natures. Objects showing both wave and particle natures are often called quantons \cite{bunge}. It was Bohr who first pointed out that both properties are mutually exclusive and formulated it as a principle of complementarity \cite{bohr}. 
 Wootters and Zurek \cite{wootters} revisited Bohr's complementarity principle from the information-theoretic approach, looking at  two-slit interference in the presence of a path detector, and found that simultaneous observation of both natures is possible with the proviso that the more you observe one, the more it will obscure the other. Later, Greenberger and Yasin \cite{greenberger} formulated a quantitative bound in terms of the predictability and fringe visibility.  The \emph{predictability} was defined as \emph{a priori} information i.e., it tells one the difference between probabilities of going through different paths. Englert \cite{englert}  proposed a stronger path quantifier which was based on \emph{a posteriori} path information acquired using a path detector, and derived a bound on the path \emph{distinguishability} and fringe visibility, ${\mathcal D}^2 + {\mathcal V}^2 \le 1$. This relation, generally called the wave particle duality relation, is understood to be a quantitative statement of Bohr's principle. Of late the concept of wave particle duality has been generalized to multipath interference \cite{3slit,cd15,nduality,bagan,roy}.

\begin{figure}
\centering
\includegraphics[width=7.0 cm]{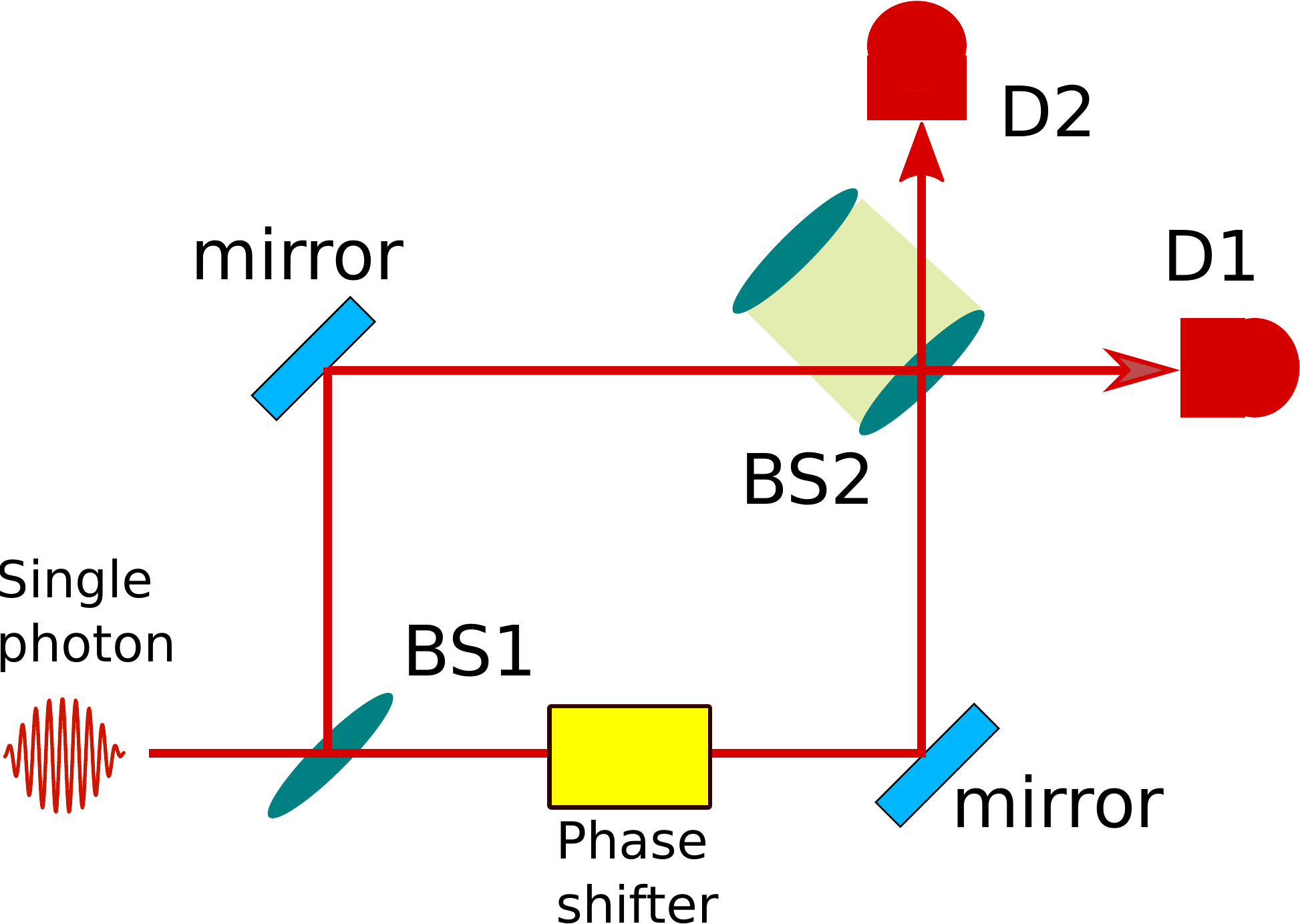}
\caption{Schematic diagram to illustrate a typical interference experiment with a quantum which-path device BS2. The beam-splitter BS2 is in a superposition of being present in the path of the photon and being away from it.}
\label{qcsetup}
\end{figure}

In a Mach-Zehnder interferometer, it is understood that in the balanced mode, only one of the detectors registers all the photons, and no photons arrive at the other detector due to destructive interference. In this situation, it is logical to believe that the photon follows both paths, which later interfere. If the second beam-splitter is removed, photons from one path can only reach a particular detector. So it is logical to assume that each photon detected by any detector came from only one path and not both. So the presence of the second beam-splitter makes the photons behave as a wave, following both paths, and in its absence they behave like particles, following only one path at a time. Wheeler introduced an idea that if the choice of removing or retaining the beam-splitter is made after the photon has traversed most of its path, one can affect the past of the particle in the sense of making sure, even after a delay, that the photons behave like a wave or like a particle \cite{wheeler}. This ``delayed choice" idea has been a subject of debate for a long time. Some years back, a proposal was made by Ionicioiu and Terno \cite{terno} suggesting that the second beam-splitter could be a quantum beam-splitter (QBS), such that it is in a quantum superposition of being present and absent (see Fig. \ref{qcsetup}). The idea was that this would force the photon to be in a superposition of wave and particle natures. This ``quantum delayed choice" experiment, with a quantum beam-splitter immediately became a subject of attention, and many experimental and theoretical studies were carried out \cite{celeri,tang,peruzzo,kaiser,tangexpt,zheng}. 

Apart from the obvious relevance of this new class of experiments to Wheeler's delayed choice idea, there have been speculations that the superposition of wave and particle natures might violate complementarity. In particular, some claims of exceeding the bound set by the two-path duality relation of the kind $\mathcal{D}^2+\mathcal{V}^2 \leq 1$ have been made \cite{tang}. In this paper, we investigate the issue of wave particle duality in the more general scenario of $n$-path interference, where the path detector is in a quantum superposition of being present and absent.

\section{Wave-particle duality in multipath interference}
\label{Preliminaries}
\subsection{Duality relation for pure quanton and quantum path detector}
\begin{figure}
\centering
\includegraphics[width=7.0 cm]{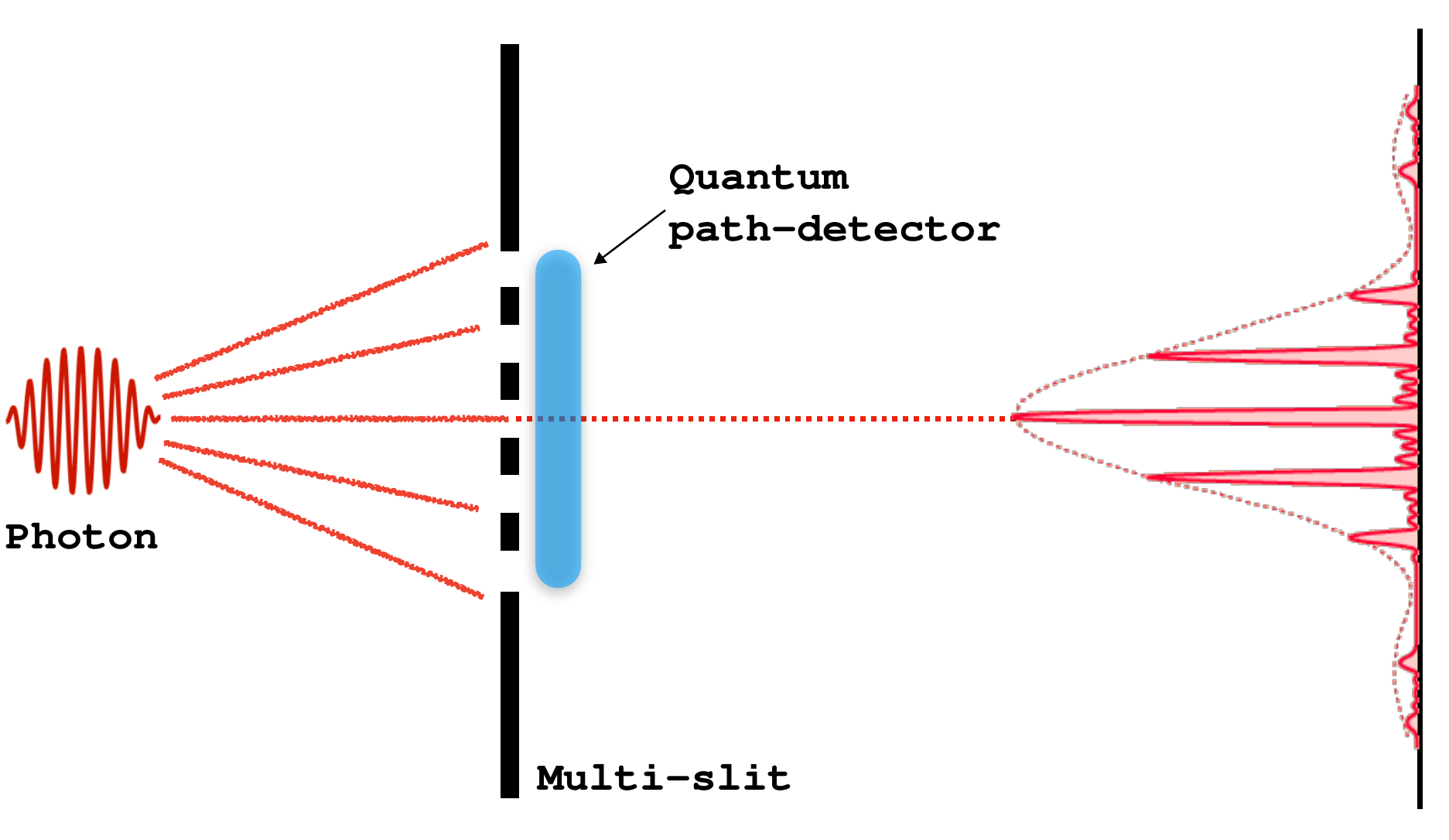}
\caption{Schematic drawing of an $n$-path interference experiment with a quantum which-path detector. The path-detector is in a superposition of being present and absent in the path of the photon.}
\label{nslit}
\end{figure}

Consider an $n$-path interference experiment (see Fig. \ref{nslit}) with pure initial quanton state
\begin{equation} 
|\psi_{in}\rangle=\sum_{i=1}^n\sqrt{p_i}\, {|\psi_i\rangle},
\end{equation}
 where ${p_i}$ is the probability of acquiring the $i$th path and $|\psi_i\rangle$ forms an orthonormal basis. 

We use a quantum path detector (QPD) to detect the path acquired by a quanton. There are two degrees of freedom associated with it. One is its location, which is assumed to have two states, $|Y\rangle$ corresponding to it being \emph{present} in the paths of the quantum and $|N\rangle$ corresponding to be being  \emph{absent} from the path. The other degree of freedom is its internal state denoted by $|d_i\rangle$, which corresponds to it detecting the path of the quanton. Initially, the QPD is assumed to be in the state $|d_0\rangle$, and if the quanton goes through the $k$th path, the QPD state changes to $|d_k\rangle$. So the full initial detector state is given by
\begin{equation}
|\phi_0\rangle = |d_0\rangle \left(c_1\, |Y\rangle
+ c_2\, |N\rangle \right) ,
\label{phi0}
\end{equation}
where ${c_1}$ is the amplitude of QPD presence and  $c_2$ the amplitude of its absence; $c_1^2+c_2^2 =1$. The state represents the QPD being in a superposition of the two locations.
%The $\theta=0$, indicate the presence of path detector in front of slit, while  $\theta=\frac{\pi}{2}$ means it is absent.

Initially, the joint state of quanton and QPD is given by 
\begin{eqnarray}\label{rhin}
|\Psi_{in}\rangle = |\psi_{in}\rangle|\phi_0\rangle 
 = \sum_{i=1}^n\sqrt{p_i}\, {|\psi_i\rangle}
	|d_0\rangle \left(c_1\, |Y\rangle + c_2\, |N\rangle \right),~~~~~
\end{eqnarray}
which denotes a pure state of the quanton with amplitude $\sqrt{p_k}$ to go
through the $k$th path, being in the state $|\psi_k\rangle$, and the QPD in
a superposition of being present and absent.
The interaction can be represented by a controlled unitary operation,
${U}$. The combined state of quanton and QPD, after the quanton has traversed
the paths and interacted with the QPD, can be written as
\begin{eqnarray}\label{rhstate}
|\Psi\rangle &=& c_1\big[\sum_{i=1}^n\sqrt{p_i}\,|\psi_i\rangle|d_i\rangle\big] |Y\rangle 
+ c_2\big[\sum_{i=1}^n\sqrt{p_i}\,|\psi_i\rangle\big] |d_0\rangle|N\rangle .\nonumber\\
\end{eqnarray}
The first term in the above equation represents the quanton states entangled
with the internal states of the QPD, when the QPD is present in the path of
the quanton, i.e., it is in the state $|Y\rangle$. Here path information of
the quanton is encoded in the $|d_i\rangle$ states of the QPD, and the quanton
behaves as a particle. The second term
represents the pure state of the quanton in a superposition of $n$ paths,
acting like a wave, and the QPD away from its path, in the state $|N\rangle$.
The state (\ref{rhstate}) can be written as $c_1|\text{particle}\rangle|Y\rangle + c_2|\text{wave}\rangle|N\rangle$, and represents a
superposition of particle nature and wave nature, with amplitudes $c_1$
and $c_2$, respectively. It is the most natural generalization of the
wave and particle superposition states studied earlier (without a QPD)
\cite{celeri,tang,peruzzo,kaiser,tangexpt,zheng},
to the case where there is a real QPD present. A similar state has also been
used in a very recent work using a QPD \cite{wang}. It may be convenient to
use the density operator formalism if one wants to generalize the analysis to
mixed states. The density operator for the state (\ref{rhstate}), is given by
\begin{equation}\label{rh}
\rho_{\text{QD}}=\sum_{i,j=1}^n\sqrt{p_i p_j}\, |\psi_i\rangle \langle \psi_j|\otimes U_i |\phi_0\rangle \langle \phi_0| U_j^\dag ,
\end{equation}
where $U_i |\phi_0\rangle= c_1\, |d_i\rangle|Y\rangle
+ c_2\, |d_0\rangle|N\rangle $.

The above interaction creates entanglement between the quanton and path detector. Thus, for gaining knowledge of the path of the quanton, it is sufficient to do a measurement on the states $|d_i\rangle$ of the QPD. Here we will use the unambiguous quantum state discrimination (UQSD) method for gaining the path information \cite{3slit,cd15}. For wave information  we will use $l_1$ norm measure of coherence  \cite{cd15,coherence,tqcoherence}.
Let us now look at the path distinguishability and the measure of coherence.

\textit{{Path distinguishability}}: 
Based on UQSD, the path-distinguishability for $n$-path interference \cite{3slit,cd15}, is given by
\begin{eqnarray}
\mathcal{D}_Q &:=& 1 - {1\over n-1}\sum_{i\neq j} \sqrt{p_i p_j}\,  |\langle \phi_0| U_j^\dag  U_i |\phi_0\rangle| \nonumber\\&=& 1-{1\over n-1}\sum_{i\neq j} \sqrt{p_i p_j}\, \left( c_1^2\, |\langle{d_j} |{d_i}\rangle|+c_2^2 \right).
\label{DQ}
\end{eqnarray}
It is essentially the maximum probability with which the states $U_i|\phi_0\rangle$ can be \emph{unambiguously} distinguished from each other.

\textit{{Quantum coherence}}: Quantum coherence \cite{coherence,cd15,tqcoherence} gives the wave nature of a quanton, given by
\begin{equation}
\label{coh}
{\mathcal C}(\rho) := {1\over n-1}\sum_{i\neq j} \abs{\rho_{ij}} ,
\end{equation}
where $n$ is the dimensionality of the Hilbert space. The reduced density matrix of the quanton can be obtained by tracing out all the  states of the QPD:
\begin{eqnarray}
\label{rdm}
\rho_Q &=& \sum_{i,j=1}^n \sqrt{p_i p_j}\, \mbox{Tr}\left( U_i |\phi_0\rangle \langle \phi_0| U_j^\dag \right)  |\psi_i\rangle\langle\psi_j|.
\end{eqnarray}

The set $\{ |\psi_i\rangle\}$ forms a complete basis for the $n$ path setup. Thus, the coherence can be obtained using the reduced density matrix
\begin{eqnarray}
\label{cohr}
{\mathcal C} &=& {1\over n-1}\sum_{i\neq j} \abs{\langle\psi_i|\rho_Q|\psi_j\rangle}
\nonumber\\
&=& {1\over n-1}\sum_{i\neq j} \sqrt{p_i p_j}\, \abs{ \mbox{Tr}\left( U_i |\phi_0\rangle \langle \phi_0| U_j^\dag \right) }. 
\end{eqnarray}
Using Eq. (\ref{phi0}), we get the following form:
\begin{eqnarray}
\label{cohfn}
{\mathcal C}={1\over n-1}\sum_{i\neq j} \sqrt{p_i p_j}\, \left( c_1^2\, |\langle{d_j} |{d_i}\rangle|+c_2^2 \right).
 \end{eqnarray}
Combining Eqs. (\ref{DQ}) and (\ref{cohfn}), we get 
\begin{equation}
{\mathcal D}_Q + {\mathcal C} = 1.
\label{duality}
\end{equation}
This is a tight wave particle duality relation which had been derived earlier for $n$-path interference \cite{cd15}. So, the relation continues to hold even in the case of a QPD.

\textit{{Two-path experiment:}}
For $n=2$ and $p_1=p_2=\tfrac{1}{2}$,  the path distinguishability (\ref{DQ}) and coherence (\ref{cohfn}) becomes
\begin{equation}
{\mathcal D}_Q = c_1^2 \left(1- \abs{\langle{d_1} |{d_2}\rangle} \right)
\end{equation}
\begin{equation}
{\mathcal C}= 1- c_1^2 + c_1^2 \abs{\langle{d_1} |{d_2}\rangle}.
\end{equation}
Our result reproduces the earlier result \cite{qtwist} for a two path experiment in the presence of a QPD, while recognizing that for two paths, the coherence $\mathcal{C}$ is identical to the traditional visibility $\mathcal{V}$ \cite{tqcoherence}. It also satisfies Eq. (\ref{duality}) in the same way.
\subsection{Superposition of wave and particle natures}
%\begin{equation}
%|\Psi\rangle=|\psi_{0}\rangle |\phi_0\rangle 
%\end{equation}

The preceding analysis is for the behavior of the quanton irrespective of the \emph{location} state of the QPD. One might argue that one would get the same result if QPD were not in the superposition state (\ref{phi0}), but in a mixed state of being present and absent. To really see the effect of the QPD being in a superposition, one should look at the behavior of the quanton conditioned on obtaining a superposition location state of the QPD. To this end, let us assume the QPD location is measured in certain basis and collapses to
\begin{equation}\label{bsp}
    |\phi_{\alpha}\rangle= \cos{\alpha}\, |Y\rangle+ \sin{\alpha}\, |N\rangle ,
\end{equation}
which is the state just for the location degree of the QPD.

The interaction can be represented by a controlled unitary operation, $\mbox{U}$. The combined state of quanton and QPD  can be written as
%\begin{equation}\label{rh}
%\rho_{qd}=\sum_{i,j=1}^n\sqrt{p_i p_j}\, |\psi_i\rangle \langle \psi_j|\otimes U_i |\phi_0\rangle \langle \phi_0| U_j^\dag.
%\end{equation}

\begin{equation}\label{rh1}
\rho_{\text{QD}}=\sum_{i,j=1}^n\sqrt{p_i p_j}\, |\psi_i\rangle \langle \psi_j|\otimes |d_i'\rangle \langle d_j'|.
\end{equation}
where $ |d_i'\rangle \equiv \langle \phi_{\alpha}| U_i |\phi_0\rangle =  c_1 \cos{\alpha}\, |d_i\rangle
+ c_2 \sin{\alpha} \, |d_0\rangle$; with normalization condition  $c_1^2 \cos^2{\alpha}+ c_2^2 \sin^2{\alpha}=1.$

%\begin{equation}
%|\psi\rangle= \sum_i \sqrt{p_i}\ \left(c_1 |d_i\rangle |Y\rangle + c_2 |\d_0\rangle |N\rangle\right)
%\end{equation}

The above interaction creates the entanglement between the quanton and path detector, with the QPD out of the picture now. Following the earlier procedure, we will use the UQSD method for gaining the path information and coherence for wave information.
Based on UQSD, the path-distinguishability for $n$-path interference is given by
\begin{eqnarray}
{\mathcal D}_Q &=& 1-\tfrac{1}{n-1}\sum_{i\neq j} \sqrt{p_i p_j}\, |\big( c_1^2 \cos^2{\alpha}\,\langle{d_j} |{d_i}\rangle+c_2^2 \sin^2{\alpha} \nonumber\\&& + \tfrac{c_1 c_2}{2} \sin{2 \alpha} \left( \langle d_j|d_0 \rangle+\langle d_0|d_i \rangle\right) \big)| .
%\\&=&1-\tfrac{1}{n-1}\sum_{i\neq j} \sqrt{p_i p_j}\, \left( 1- c_1^2 \cos^2{\alpha}(1-\abs{\langle{d_j} |{d_i}\rangle})\right).\nonumber\\
\label{DQ1}
\end{eqnarray}
The reduced density matrix of the quanton can be obtained by tracing out the detector states
\begin{eqnarray}
\label{rdm1}
\rho_Q &=& \sum_{i,j=1}^n \sqrt{p_i p_j}\, \mbox{Tr}\left(|d_i'\rangle|\langle d_j'| \right)  |\psi_i\rangle\langle\psi_j|.
\end{eqnarray}
The set $\{ |\psi_i\rangle\}$ forms a complete incoherent basis for $n$ path setup. Thus, the coherence can be obtained using the reduced density matrix
\begin{eqnarray}
\label{cohr1}
{\mathcal C} %&=& \tfrac{1}{n-1}\sum_{i\neq j} \abs{\langle\psi_i|\rho_Q|\psi_j\rangle}
%\nonumber\\
&=& \tfrac{1}{n-1}\sum_{i\neq j} \sqrt{p_i p_j}\,  \abs{\langle d_j'|d_i'\rangle}. 
\end{eqnarray}
Using Eq. (\ref{phi0}), we get the following form:
\begin{eqnarray}
\label{cohf1}
{\mathcal C}&=&\tfrac{1}{n-1}\sum_{i\neq j} \sqrt{p_i p_j}\, |\big( c_1^2 \cos^2{\alpha}\,\langle{d_j} |{d_i}\rangle+c_2^2 \sin^2{\alpha} \nonumber\\&& + \tfrac{c_1 c_2}{2} \sin{2 \alpha} \left( \langle d_j|d_0 \rangle+\langle d_0|d_i \rangle\right) \big)|.
%\left( 1- c_1^2 \cos^2{\alpha}(1-\abs{\langle{d_j} |{d_i}\rangle})\right).
 \end{eqnarray}
Combining Eqs. (\ref{DQ1}) and (\ref{cohf1}), we get 
 \begin{equation}
 {\mathcal D}_Q + {\mathcal C} = 1.
 \label{duality1}
 \end{equation}
Thus, even when quanton is forced to be in a superposition of wave and particle
natures, the usual wave-particle duality relation continues to hold. This is at variance with earlier claims suggesting that wave-particle duality relations
are violated in such a situation. 

\subsection{Perspectives}

At this stage, it may be useful to analyze these results in light of various earlier works. It is widely believed that the superposition of wave and particle natures may lead to a violation of the complementarity. However, most experiments that have been carried out, do not involve a path-detecting device. Rather, the beam-splitter BS2 (see Fig. \ref{qcsetup}) is in a superposition of being present and absent. So, in the situation where BS2 is in a superposition, there is no way of knowing if a particular photon received at (say) D1, followed one path or both paths. In such a situation, one can only talk of the probability of taking one path or the other; the duality relation that is meaningful is the one derived by Greenberger and Yasin \cite{greenberger}. The duality relation pertaining to \emph{detecting} which path the quanton followed, derived by Englert \cite{englert}, is not applicable in such scenarios. 

The analysis carried out in the previous subsections shows that complementarity is always respected in the multipath interference experiment which has a path-detecting device in the superposition of being present and absent. Equation (\ref{DQ}) has a nice interpretation that the path-detecting states $|d_i\rangle$ are present with a probability $c_1^2$ and absent with probability $c_2^2$. And it leads to the perfect duality relation (\ref{duality}). However, if one naively uses the same definition, which appears reasonable, for the case where the quanton is really forced to be in a superposition of wave and particle behaviors, one will run into a problem. With that reasoning, one would imagine that the path-detecting states $|d_i\rangle$ are present with a probability $c_1^2\cos^2\alpha$ and absent with probability probability $c_2^2\sin^2\alpha$. The distinguishability will then come out to be ${\mathcal D}_Q = 1-\tfrac{1}{n-1}\sum_{i\neq j} \sqrt{p_i p_j}\, |\big( c_1^2 \cos^2{\alpha}\,\langle{d_j} |{d_i}\rangle+c_2^2 \sin^2{\alpha})|$. But the coherence in this situation will be ${\mathcal C}=\tfrac{1}{n-1}\sum_{i\neq j} \sqrt{p_i p_j}\, |\big( c_1^2 \cos^2{\alpha}\,\langle{d_j} |{d_i}\rangle+c_2^2 \sin^2{\alpha} + \tfrac{c_1 c_2}{2} \sin{2 \alpha} \left( \langle d_j|d_0 \rangle+\langle d_0|d_i \rangle\right) \big)|$. It is easy to see that the sum $\mathcal{D}_Q+\mathcal{C}$ may exceed $1$ because of the term $\tfrac{c_1 c_2}{2} \sin{2 \alpha} (\langle d_j|d_0 \rangle+\langle d_0|d_i \rangle)$, which is a signature of interference between the wave and particle natures. One may naively interpret it as a violation of complementarity. 
However, recognizing that the paths of the quanton are correlated with $ |d_i'\rangle \equiv \langle \phi_{\alpha}| U_i |\phi_0\rangle =  c_1 \cos{\alpha}\, |d_i\rangle
+ c_2 \sin{\alpha} \, |d_0\rangle$, and not just with $|d_i\rangle$, one can see that the unambiguous discrimination of $|d_i'\rangle$ is what will yield the correct distinguishability (\ref{DQ1}). This distinguishability leads to the correct duality relation (\ref{duality1}).

So we see that even in the scenario where there is an interference between the wave and particle natures, quantum complementarity is fully respected, governed by the wave particle duality relation (\ref{duality1}). In the experiments where there is no real path-detector in place, it is all the more likely to come to an erroneous conclusion regarding the violation of complementarity.

\subsection{Generalized duality relation}
We extend our analysis for a noisy scenario. The mixed quanton state can be taken as $\rho_{in}=\sum_{ij} \rho_{ij}|\psi_i \rangle \langle \psi_j|$. The initial joint state of a quanton and a detector system can be written as $\rho'^{\rm(in)}_{\text{QD}}=\rho_{in}\otimes \rho^{(0)}_\phi$. 
The effect of noise on the QPD can be represented as
\begin{equation}
\label{rhon}
\rho_\phi^{(0)} \longrightarrow \widetilde\rho_\phi^{(0)} =\sum_{i}K_{i}\rho_\phi^{(0)} K^{\dagger }_{i},
\end{equation}
with completeness relation $\sum_{i}K^{\dagger }_{i}K_{i}=\mathcal{I}$. The spectral decomposition of the transformed QPD  can then be written as
\begin{equation}
\label{spectral}
\widetilde\rho_\phi^{(0)} =\sum_{k} {r_{k}} |\phi_k \rangle \langle \phi_k|,
\end{equation}  
where  $\sum_{k} r_k = 1$, $r_k\ge0$, and $ \langle \phi_k|\phi_l\rangle =\delta_{kl}$.

 The combined quanton-QPD state, when QPD is considered in state Eq. (\ref{bsp}), can be written as

\begin{equation}
\label{rhoqd}
\rho_{\text{QD}}'=\sum_{i,j=1}^n \rho_{ij} |\psi_i\rangle \langle \psi_j|\otimes \sum_{k} {r_{k}}  |d'_{ki}\rangle \langle d'_{kj}|\,
\end{equation}
where $ |d_{ki}'\rangle \equiv \langle \phi_{\alpha}| U_i |\phi_k\rangle =  c_1 \cos{\alpha}\, |d_{ki}\rangle +c_2 \sin{\alpha} |d_k\rangle$
 The path distinguishability for mixed QPD (\ref{spectral}) can be calculated using
\begin{equation}
\mathcal{D}_Q^\prime =1-\frac{1}{n-1} \sum_k r_k \sum_{i\neq j} \sqrt{\rho_{ii}\rho_{jj}} |\langle d'_{kj}  | d'_{ki} \rangle |.
\label{mixD}
\end{equation}

To find the measure of coherence, let us first calculate the reduced density matrix of the quanton, given by
\begin{eqnarray}
\label{6}
	\rho_Q' &=& \sum_{i,j=1}^n \rho_{ij} \mbox{Tr}\left( \sum_k r_k  |d'_{ki}\rangle \langle d'_{kj}| \right) |\psi_i\rangle\langle\psi_j|.
\end{eqnarray}
The  coherence comes out to be
%\begin{eqnarray}
%\label{7a}
%{\mathcal C}'= \tfrac{1}{n-1} \sum_{i\neq j}  \ \abs{ \rho_{ij} \mbox{Tr}\left( U_i^\dag %\widetilde\rho^{(0)}_\phi U_j \right) }, 
%\end{eqnarray}
%Using Eq. (\ref{spectral}), we get the following form:
\begin{eqnarray}
\label{mixC}
{\mathcal C}'&=&\tfrac{1}{n-1} \sum_{i\neq j} \left|\rho_{ij} \ \sum_k r_k \langle d'_{kj} | d'_{ki} \rangle \right| \nonumber \\ &\leqslant& \tfrac{1}{n-1} \sum_k r_k \sum_{i\neq j}  |\rho_{ij}| |\langle d'_{kj}  | d'_{ki} \rangle| .
\end{eqnarray}
 %where  $|\phi_{ki} \rangle= \cos{\theta}\, |d_{ki}\rangle|Y\rangle
 %+ \sin{\theta}\, |d_k\rangle|n\rangle $
Combining Eq. (\ref{mixD}) and Eq. (\ref{mixC}), we get
\begin{equation}
{\mathcal D}_Q'+{\mathcal C}' + \tfrac{1}{n-1} \sum_k r_k \sum_{i\neq j}
(\sqrt{\rho_{ii}\rho_{jj}} - |\rho_{ij}|)|\langle d'_{kj}|d'_{ki}\rangle| = 1.
\label{cdq}
\end{equation}
Every principal 2x2 sub matrix of (\ref{rhoqd}) is positive
semi-definite \cite{horn}, thus we have
\begin{equation}
\sqrt{\rho_{ii}\rho_{jj}} - |\rho_{ij}| \ge 0.
\label{eq:se}
\end{equation}
Therefore, we find that Eq. (\ref{cdq}) reduces to
\begin{equation}
{\mathcal D}_Q'+{\mathcal C}'  \le 1,
\label{rel}
\end{equation}
where the inequality is saturated for pure initial quanton states.

\section{Are experiments with a quantum device really unique?}

Two-path interference experiments with a quantum device have attracted lots of attention. But are these experiments really unique? In this section, we try to answer this question. 

Let us consider the setup shown in Fig. \ref{qcsetup}. Since it does not use a path-detector, the duality relations derived in the previous section are not directly applicable here. For simplicity, let us consider the QBS to be in an equal superposition state $|\phi\rangle = \tfrac{1}{\sqrt{2}} (|Y\rangle + |N\rangle)$, $|Y\rangle$ represents the situation when BS2 is in the path, and $|N\rangle$ when it is not. Let the quanton in the two paths also be in an equal superposition state $|\psi\rangle = \tfrac{1}{\sqrt{2}} (e^{i\theta}|\psi_1\rangle + |\psi_2\rangle)$, $\theta$ being an arbitrary phase difference between the two paths. The effect of BS2 is to take $|\psi_1\rangle,|\psi_2\rangle$ to $|D_1\rangle, |D_2\rangle$, the detector states of the two detectors $D_1$ and $D_2$, respectively. The transformation can be written as $U_Y|\psi_1\rangle = \tfrac{1}{\sqrt{2}} (|D_1\rangle + |D_2\rangle)$ and $U_Y|\psi_2\rangle = \tfrac{1}{\sqrt{2}} (|D_1\rangle - |D_2\rangle)$. If BS2 is absent, the transformation is as follows: $U_N|\psi_1\rangle = |D_2\rangle$ and $U_N|\psi_2\rangle = |D_1\rangle$. The action of the QBS can be represented by a unitary operator $U_{\text{QBS}} = U_Y\otimes|Y\rangle\langle Y| + U_N\otimes|N\rangle\langle N|$.  Using this, the effect of the QBS on the quanton can be written as follows:
\begin{eqnarray}
U_{\text{QBS}}|\psi\rangle\otimes|\phi\rangle &=& \tfrac{1}{2}\Big[(U_Y(e^{i\theta}|\psi_1\rangle+|\psi_2\rangle)|Y\rangle\nonumber\\
&& + U_N(e^{i\theta}|\psi_1\rangle+|\psi_2\rangle)|N\rangle\Big] \nonumber\\
&=& \Big(\tfrac{|N\rangle}{2}+e^{\tfrac{i\theta}{2}}\cos\tfrac{\theta}{2}\tfrac{|Y\rangle}{\sqrt{2}}\Big)|D_1\rangle \nonumber\\
&& + e^{\tfrac{i\theta}{2}}\Big(e^{\tfrac{i\theta}{2}}\tfrac{|N\rangle}{2}+i\sin\tfrac{\theta}{2}\tfrac{|Y\rangle}{\sqrt{2}}\Big)|D_2\rangle~~~~~~
\label{qbs}
\end{eqnarray}
The above relation implies that detectors $D_1$ and $D_2$ click with
probabilities $\tfrac{1}{2}+\tfrac{1}{4}\cos\theta$ and $\tfrac{1}{2}-\tfrac{1}{4}\cos\theta$, respectively.

Let us consider a setup similar to the one shown in Fig. \ref{qcsetup},
except that the second beam-splitter BS2 is not a quantum device but a
classical \emph{biased beam-splitter} with reflection and transmission
coefficients given by $|r|^2$ and $|t|^2$, respectively, such that
$|r|^2+|t|^2=1$. The action of a biased beam-splitter can be described by the operator $U_{\text{BBS}}=(r|D_1\rangle + t|D_2\rangle)\langle\psi_1|
+ (t|D_1\rangle - r|D_2\rangle)\langle\psi_2|$. It transforms the incoming state $|\psi\rangle$ as
\begin{eqnarray}
U_{\text{BBS}}|\psi\rangle &=& \tfrac{1}{\sqrt{2}}\Big[(e^{i\theta}r+t)|D_1\rangle
+ (e^{i\theta}t-r)|D_2\rangle\Big] .
\label{bbs}
\end{eqnarray}
One can verify that if $\theta=0$ and $r=t=\tfrac{1}{\sqrt{2}}$, the
quanton will always land at the detector $D_1$. The state (\ref{bbs}) implies
that detectors $D_1$ and $D_2$ click with probabilities
$\tfrac{1}{2}+rt\cos\theta$ and $\tfrac{1}{2}-rt\cos\theta$, respectively.
For $rt=\tfrac{1}{4}$, one cannot experimentally distinguish between this
situation and the previous one, described by (\ref{qbs}), involving a
QBS.
The original proposal claimed that one can correlate the detected quantons with the $|Y\rangle$ and $|N\rangle$ states, and get wave or particle natures \cite{terno}. But even in doing that, at a time one can see either wave nature or particle nature. A similar effect can be achieved by randomly removing BS2 from the quanton path. 

Recognizing the fact that correlating with $|Y\rangle$ and $|N\rangle$ states was like a statistical effect, some authors referred to it as a \emph{classical 
mixture} of wave and particle natures, and suggested that to get
a true superposition, the quanton be observed conditioned on detection
of the state $|\phi_{\alpha}\rangle = \cos\alpha|Y\rangle + \sin\alpha|N\rangle$ \cite{tang,kaiser,zheng}. For the interesting case of $\alpha=\pi/4$, the
(unnormalized) state of the quanton in that situation will be 
\begin{eqnarray}
\langle\phi_{\alpha}|U_{\text{QBS}}|\psi\rangle &=& \tfrac{1}{2}\Big(\tfrac{1}{\sqrt{2}}+e^{\tfrac{i\theta}{2}}\cos\tfrac{\theta}{2}\Big)|D_1\rangle \nonumber\\
&& + \tfrac{1}{2}e^{\tfrac{i\theta}{2}}\Big(e^{\tfrac{i\theta}{2}}\tfrac{1}{\sqrt{2}}+i\sin\tfrac{\theta}{2}\Big)|D_2\rangle .
\label{qbs1}
\end{eqnarray}
This state is indeed different from (\ref{qbs}), and the two will yield
different results. However, the state for a \emph{classical} biased 
beam-splitter, given by (\ref{bbs}), may be rewritten as
\begin{eqnarray}
U_{\text{BBS}}|\psi\rangle &=& \sqrt{2}r\Big(\tfrac{t-r}{2r}+e^{\tfrac{i\theta}{2}}\cos\tfrac{\theta}{2}\Big)|D_1\rangle \nonumber\\
&& + \sqrt{2}re^{\tfrac{i\theta}{2}}\Big(e^{\tfrac{i\theta}{2}}\tfrac{t-r}{2r}+i\sin\tfrac{\theta}{2}\Big)|D_2\rangle .
\label{bbs1}
\end{eqnarray}
For $\tfrac{t-r}{\sqrt{2}r}=1$, (\ref{bbs1}) is very similar in form
to (\ref{qbs1}), and the probability of (say) $D_2$ clicking will show the
same behavior with respect to the phase $\theta$.

The message from the preceding analysis is that the quantum
case of the QBS is different from the classical mixture case of the QBS, as has been experimentally observed earlier \cite{tangexpt}. However, both these situations can also be mimicked by an appropriately biased \emph{classical} beam-splitter. We feel it will be interesting to explore the implications of the connection between a QBS and a biased classical beam-splitter.

What about a two-path interference experiment with a real two-state path-detecting device, which is in a superposition of being present and absent, one may ask. In the following, we will show even this experiment is completely equivalent to a two-path interference experiment with a real two-state path-detecting device, which is \emph{always present}, and is not in a superposition in the sense that is being discussed here. Let us consider a two-path interference experiment with a which-way detector whose two states that correlate with the two paths of the quanton are \emph{not} orthogonal to each other. The state of the quanton and path-detector may be written as
\begin{equation}
    |\Psi\rangle = \tfrac{1}{\sqrt{2}} (|\psi_1\rangle|d_1\rangle + |\psi_2\rangle|d_2\rangle),
\label{2state}
\end{equation}
where $\langle d_1|d_2\rangle \neq 0$. Now it can be shown that the states $|d_1\rangle,|d_2\rangle$ can be represented in terms of an expanded Hilbert space as follows \cite{awpd,neha}:
\begin{eqnarray}
|d_1\rangle = \gamma|q_1\rangle + \beta|q_3\rangle,\hskip 5mm
|d_2\rangle = \gamma|q_2\rangle + \beta|q_3\rangle ,
\label{d1d2}
\end{eqnarray}
where $|q_1\rangle, |q_2\rangle, |q_3\rangle$ are orthonormal states, and $\gamma,\beta$ are certain constants which we need not specify for the present purpose. In this basis, the entangled state (\ref{2state}) has the following form
\begin{eqnarray}
|\Psi\rangle = \tfrac{1}{\sqrt{2}} \gamma[|\psi_1\rangle|q_1\rangle
+ |\psi_2\rangle|q_2\rangle] 
 + \tfrac{1}{\sqrt{2}}\beta[|\psi_1\rangle + |\psi_2\rangle] |q_3\rangle.\nonumber\\
\label{2staten}
\end{eqnarray}
This state can be interpreted as a representation of a superposition of wave and particle natures. The quanton state correlated with $|q_3\rangle$ represents a quanton showing wave nature, and the rest showing particle nature. If one were to measure an observable $Q$ which has $|q_1\rangle, |q_2\rangle, |q_3\rangle$ as three eigenstates with distinct eigenvalues, the quantons detected in coincidence with $|q_3\rangle$ will show full interference, and those detected in coincidence with $|q_1\rangle, |q_2\rangle$ will show full particle nature. This state will show all the features that the state (\ref{rh}) can show, although it involves only a conventional path detector and not a quantum device. Such a state can also be produced without expanding the Hilbert space, but by introducing a two-state ancilla system interacting with the path-detector \cite{qwpd}.

From this analysis, we conclude that although a lot of research interest was generated by the interference experiments with a quantum device, the effect they show can also be seen in conventional interference experiments. 

\section{Conclusions}

In conclusion, we have theoretically analyzed an $n$-path interference experiment where the path-detector is assumed to exist in a superposition of being present and absent from the interference path. We have shown that the $n$-path wave particle duality relation derived earlier \cite{cd15} continues to hold even in this case. The duality relation remains tight even in the situation where there is expected to be interference between the wave and particle natures of the quanton. So, the various interference experiments, with a quantum device, may be of interest for various reasons but are completely within the realm of complementarity. We have also shown that the effects seen due to a path detector in a quantum superposition, can also be seen in interference experiments with a conventional which-way detector. The effects seen in the quantum delayed choice experiment, i.e., without a real path detector, but with a QBS, can also be seen in a conventional Mach-Zehnder setup with a biased beam-splitter. 

\begin{acknowledgements}
M.A.S. acknowledges the National Key R$\&$D  Program  of  China,  Grant  No. 2018YFA0306703.
\end{acknowledgements}

\end{document}